\documentclass[journal]{IEEEtran}

\usepackage{graphicx}
\usepackage{booktabs}
\usepackage{array}
\usepackage{subcaption}
\usepackage{empheq}
\usepackage{multirow} 
\usepackage{makecell}
\newcolumntype{C}[1]{>{\centering\arraybackslash}p{#1}}
\usepackage{enumitem}
\usepackage{lipsum}
\usepackage{url}

\begin{document}

\title{A Novel Segment-Based Tracking Algorithm for \\ HLT under High-Occupancy and Complex \\ Conditions}

\author{Pengkun~Jia,
        Zhujun~Fang,
        Hang~Zhou,
        Yuhe~Huang,
        Changqing~Feng,
        Jianbei~Liu
\thanks{Manuscript received January XX, 2026. This work was supported by the National Natural Science Foundation of China under Grant No. 12575206, No. 12341503, and No. 12125505, the Hefei Comprehensive National Science Center on the STCF key technology research project, the international partnership program of the Chinese Academy of Sciences Grant No. 211134KYSB20200057, and the National Key R\&D Program of China under Contract No. 2022YFA1602200. (Corresponding author: Zhujun Fang.)}
\thanks{Pengkun Jia, Zhujun Fang, Hang Zhou, Yuhe Huang, Changqing Feng, Jianbei Liu are with the State Key Laboratory of Particle Detection and Electronics, University of Science and Technology of China, Hefei 230026, China, and also with the Department of Modern Physics, University of Science and Technology of China, Hefei 230026, China (e-mail: jpk0113@mail.ustc.edu.cn; fzj19@ustc.edu.cn; zh25@mail.ustc.edu.cn; hyh2827@mail.ustc.edu.cn; fengcq@ustc.edu.cn; liujianb@ustc.edu.cn).}}

\markboth{IEEE TRANSACTIONS ON NUCLEAR SCIENCE}%
{Shell \MakeLowercase{\textit{et al.}}: Bare Demo of IEEEtran.cls for IEEE Journals}

\maketitle

\begin{abstract}
In the High-Level Trigger (HLT) of both electron–positron and hadron collision experiments, the tracking process for large-volume gaseous detectors typically consumes a latency of hundreds of milliseconds. Upgrades of existing experiments and the development of next-generation facilities demand enhanced HLT tracking performance: handling higher detector occupancy and suppressing latency. To address high occupancy conditions, a novel HLT tracking algorithm based on track segments is proposed. This method involves constructing a pattern bank comprising 11 pre-defined patterns, optimizing edge-matrix formation using position, momentum, and timing criteria, and merging stereo superlayer segments to improve track consistency. These measures significantly reduce the number of stored segments and the size of the edge matrix, thereby lowering the complexity of global tracking. Even at 25\% occupancy, the number of elements for global tracking is reduced to approximately 400-500, while the density of the edge matrix remains below 1\%. With the depth-first search within connected components, the simulation results show that the algorithm maintains stable performance with occupancy ranging from 5\% to 25\%, achieving a data compression ratio of approximately 50\% to 70\%. Validation against the STCF offline reconstruction algorithm confirms that the HLT algorithm preserves high signal-hit retention without introducing significant adverse effects on offline tracking efficiency or on the reconstruction. These results demonstrate that the proposed algorithm can retain high-quality signal hits across a broad range of occupancy levels, indicating a strong potential for further development and adaptation to even more challenging, high-luminosity experimental conditions.
\end{abstract}

\begin{IEEEkeywords}
High level trigger, tracking, track segment, high occupancy. 
\end{IEEEkeywords}

\IEEEpeerreviewmaketitle

\section{INTRODUCTION}

\IEEEPARstart{T}{he} trigger system is an indispensable component of the detector spectrometer in both hadron or electron-positron collision experiments \cite{cite1}, with a primary task of achieving real-time online event selection, eliminating background data slice and hits, and minimizing the data size for storage. In both the classic trigger architecture and the “trigger less” architecture, the most complex algorithms are implemented at the high-level trigger (HLT) level. The HLT faces a trade-off between latency and performance. In HLT, the tracking algorithm is used to achieve high track-finding efficiency to facilitate subsequent accurate reconstruction, and this algorithm takes a large proportion ratio in latency, typically in hundreds of milliseconds, such as 200-400 ms in ATLAS \cite{cite2}, 451 ms in CMS \cite{cite3}, more than 300 ms in Belle II \cite{cite4}. For the tracking system based on large gaseous detector, such as drift chamber and time projection chamber (TPC), more signal hits in tracks generating a much more complex tracking environment. Thus, the HLT tracking algorithm faces more pressure in the trade-off between latency and performance.

The upgrades of existing experiments like LHC and Belle II \cite{cite5, cite6, cite7}, as well as next-generation collider experiments such as FCC-ee \cite{cite8}, CEPC \cite{cite9}, and super tau-charm facility (STCF) \cite{cite10}, aim to achieve higher luminosity and higher event rates. Thus, the increasing occupancy in tracking detector and the much higher data size press significant pressure on the HLT tracking algorithm. The occupancy refers to the ratio of output channels in the detector generating hit signals to the total number of channels, with contributions from both physical collision events and backgrounds. In hadron collisions, a large number of uninteresting physical collision events are produced, leading to a significant increase in detector occupancy as luminosity rises. Similarly, in electron-positron collisions, beam-induced backgrounds and luminosity-related backgrounds grow with higher luminosity, also contributing to increased detector occupancy. Some research indicates that the luminosity upgrades in the new experiments will lead to a data size increase by several times or even one of magnitude. In the Phase-II upgrade in ATLAS, the data size is expected to rise from 2.9 MB/event to an estimated 5.2 MB/event, and almost all of this additional data originates from the tracking detector system \cite{cite11}. After the ALICE RUN3 upgrade, the occupancy of the TPC increased by approximately five times, resulting in the estimation of a doubling of the fake track rate for tracks in the 100 MeV–GeV momentum range when the Pb-Pb collision frequency increased from 10 kHz to 50 kHz \cite{cite12}.

Several HLT tracking algorithms are developed in the current experiments, such as the Hough transform and cellular automaton. Typically, the classic conventional tracking algorithms for drift chamber can maintain stable and satisfactory performance up to approximately 15\% occupancy \cite{cite13}. However, when confronted with high luminosity, high occupancy, and large input data volumes, these algorithms also exhibit some problems. In Belle II, when the luminosity increases to $4.7 \times 10^{34} cm^{-2}s^{-1}$ , the long HLT latency is pushing the online data processing capacity of current computing resources close to its bottleneck. Thus, the even larger data volumes from further luminosity increases will pose a significant challenge \cite{cite14}. To solve this problem, Belle II is investigating new HLT tracking algorithms as potential replacements, with the approaches such as neural networks and central drift chamber driven tracking algorithms (CAT) \cite{cite13}. The HLT tracking algorithms based on cellular automaton and Kalman filter are developed for the large radial TPC in ALICE. After GPU parallel acceleration, the HLT track-finding time per event can be reduced by approximately 24\%, reaching 110 ms \cite{cite15}. However, the cellular automaton algorithm performs track finding layer by layer, and updating the state of each cell requires calculations based on the states of its neighboring cells.

The tracking logic based on track segments is commonly used in trigger systems, particularly at the L1 trigger. With the support of modern FPGAs, microsecond-level tracking can be achieved even under high data input. The segment-based tracking methods used in L1 trigger primarily rely on the positional continuity of segments, while with the limited use of the momentum continuity inherent in these segments. In HLT, the momentum continuity of the track segments is the key parameter to improving the tracking performance in complex environments \cite{cite13}. However, the form and usage of this parameter directly affect algorithm complexity, performance, and latency, making it worthwhile to conduct in-depth optimization.

In this paper, a novel segment-based tracking method designed for HLT is proposed and described. This method employs multi-dimensional filtering of segment associations by simultaneously leveraging positional, momentum, and timing continuity across segments. This approach substantially reduces the candidate segment pool for downstream processing and accelerates the subsequent global tracking stage. The proposed algorithm is designed for even higher occupancy, such as 25\%. Furthermore, the method integrates dedicated optimizations for challenging tracking cases including ultra-low transverse momentum tracks, looped tracks, crossing tracks, displaced tracks, and cluster-induced backgrounds, with the goal of improving overall tracking robustness and adaptability under high-occupancy conditions

\section{ALGORITHM FRAMEWORK DESIGN}

\begin{figure}[!t]
\centering
\includegraphics[width=\linewidth]{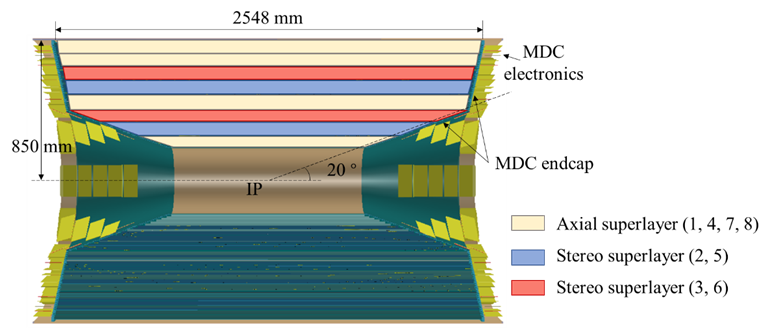} 
\caption{Geometry and superlayer layout of the STCF MDC conceptual design.} 
\label{fig:1}
\end{figure}

The design of the HLT tracking algorithm is closely tied to the structural layout of the tracking system. In this work, the algorithm is developed based on the main drift chamber (MDC) of the STCF, following its conceptual design \cite{cite10}. As illustrated in Figure~\ref{fig:1}, the STCF MDC contains 48 cell layers organized into eight superlayers, each comprising six layers. From innermost to outermost, superlayer 1, 4, 7, 8 are axial, while the 2-3, 5-6 constitute two pairs of stereo superlayers.

\begin{figure}[!t]
\centering
\includegraphics[width=\linewidth]{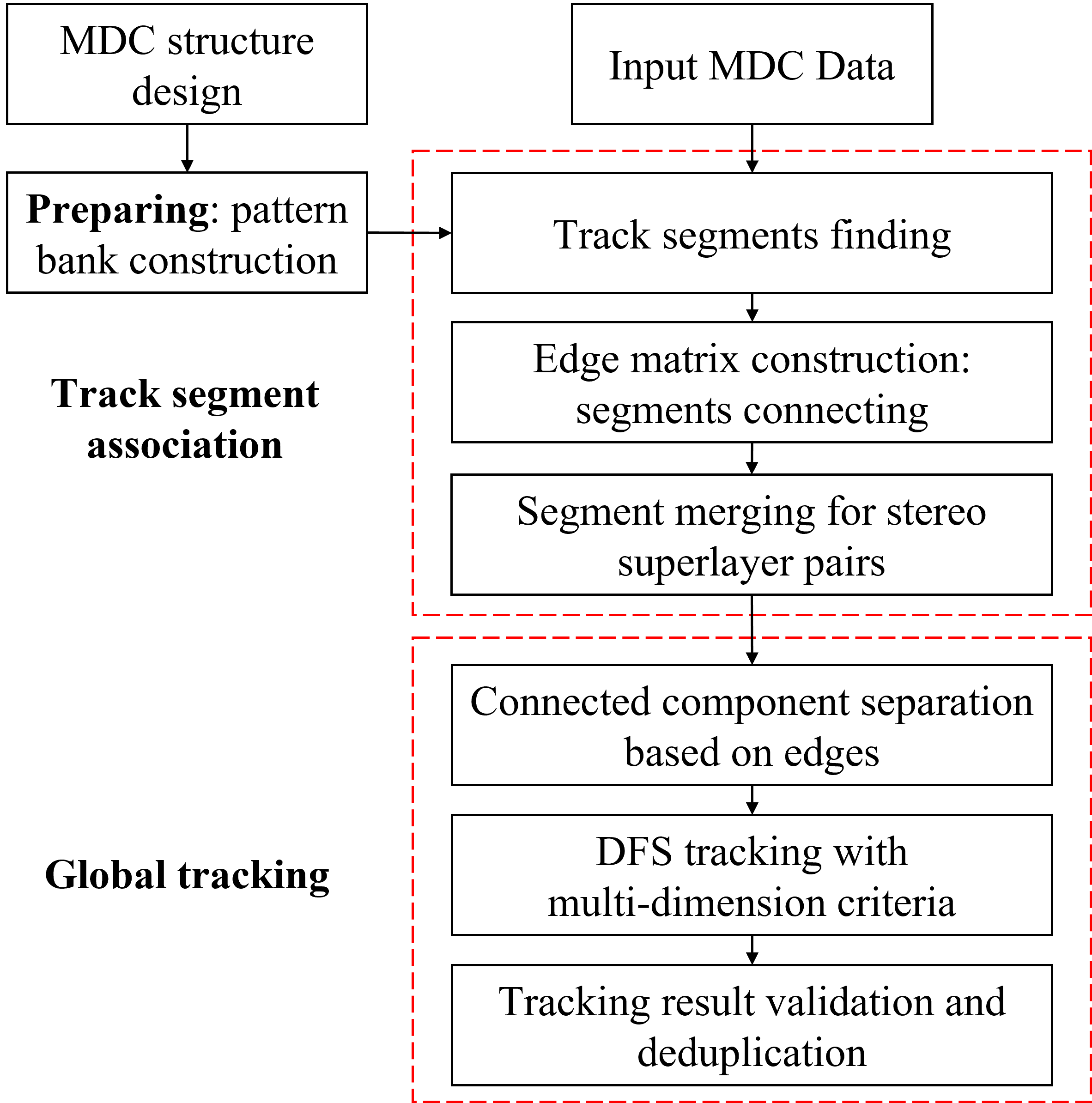} 
\caption{The Flowchart of the segment-based HLT tracking algorithm.} 
\label{fig:2} 
\end{figure}

Figure~\ref{fig:2} presents the flowchart of the segment-based HLT tracking algorithm. The algorithm comprises three primary stages: a preparatory stage, a track segment association stage, and a global tracking stage. In the preparatory stage, a pattern bank is constructed to encapsulate highly compressed, low-complexity typical segment patterns for fast matching. The track segment association stage consists of three sequential steps: first, identifying track segments within each superlayer. Second, constructing an edge matrix that records potential connections between these segments. Third, merging the stereo superlayer segment pairs into one segment. The global tracking stage can be realized through various methods. In this work, the connected component analysis enhanced by a depth-first search (DFS) is employed. This DFS utilizes multi-dimensional criteria to form complete track candidates from each seed segment. Finally, a deduplication process is applied to filter out low-quality or duplicated tracks from the results.

As mentioned before, the HLT tracking algorithm must handle complex input data with higher occupancy, such as the crossing tracks, looped tracks, and the track with ultra-low transverse momentum. Furthermore, the finite detection efficiency of the MDC cells must be accounted for, as it can result in incomplete track segments. For example, when the MDC cell detection efficiency is 95\%, approximately 3\% of directly penetrating segments will have only four hits in a six-layer superlayer. To achieve optimal tracking and data compression performance under these conditions, several critical components of the segment-based algorithm should be carefully developed. These include:
\begin{enumerate}[label=\arabic*)]
    \item the construction of an efficient pattern bank
    \item the computation of a low-complexity edge matrix
    \item the merging of stereo superlayer segments
    \item the design of multi-dimensional criteria for DFS tracking
\end{enumerate}

Addressing these components effectively is key to handling the challenging conditions in HLT tracking. Their detailed implementation is described in the subsequent sections.

\section{TRACK SEGMENT PATTERNING, FINDING AND ASSOCIATING}

\subsection{Track segment reconstruction: from pattern construction to track association}

\begin{figure}[!t]
\centering
\includegraphics[width=\linewidth]{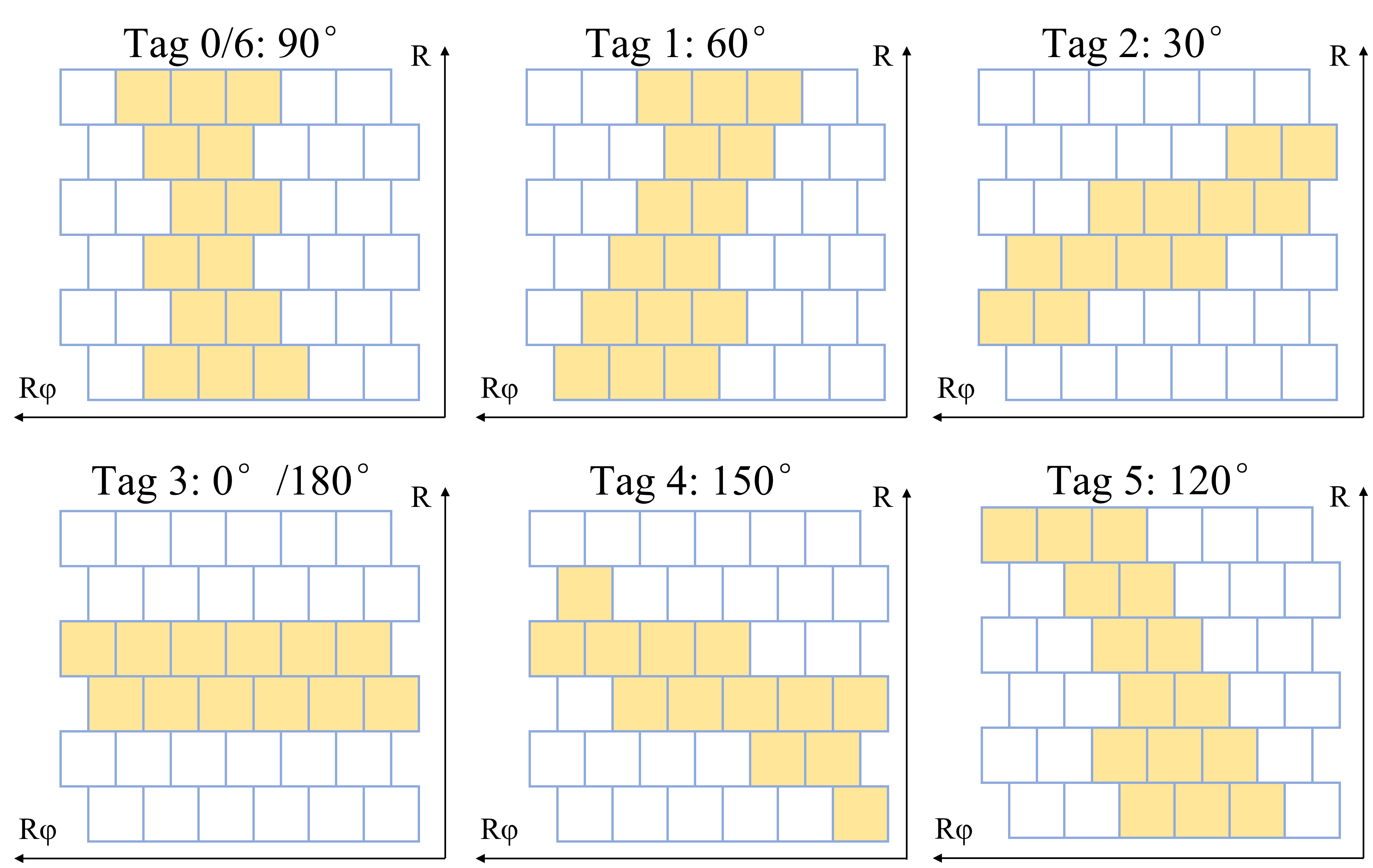} 
\caption{The definition of six kinds of tags for segment patterns.}
\label{fig:3}
\end{figure}

In HLT tracking algorithm, enforcing the momentum continuity criterion along track segments is crucial for suppressing the false tracking rate. The segment momentum information is typically derived from the curvature, which is geometrically represented by the angle between the segment’s tangent and the radial line from the interaction point (IP) to the segment. This angle serves as a fundamental parameter in the tracking logic \cite{cite13}. In this proposed HLT tracking algorithm, the momentum information of the segment is extracted and get assigned as a discrete tag, which serves as the primary feature for the segment pattern design. Figure~\ref{fig:3} defines the complete set of segment momentum tags. Within this tagging scheme, valid tracks must correspond to one of two monotonic sequences according to the charge: $0 \rightarrow 1 \rightarrow 2 \rightarrow 3 \rightarrow 4 \rightarrow 5$ or $0(6) \rightarrow 5 \rightarrow 4 \rightarrow 3 \rightarrow 2 \rightarrow 1$. Furthermore, offline analysis of looped tracks has shown that the first half-turn provides the most reliable curvature measurement, while subsequent turns introduce confusion and noise. Therefore, the HLT algorithm is designed to retain only hits from the first half of the initial turn. Consequently, the valid monotonic progression is simplified to $0 \rightarrow 1 \rightarrow 2 \rightarrow 3$ or $0(6) \rightarrow 5 \rightarrow 4 \rightarrow 3$. By enforcing this simplified monotonic rule, the number of combinatorial possibilities for connecting segments is greatly suppressed. This results in significant conservation of computing resources and a reduction in latency.

\begin{figure}[!t]
\centering
\includegraphics[width=\linewidth]{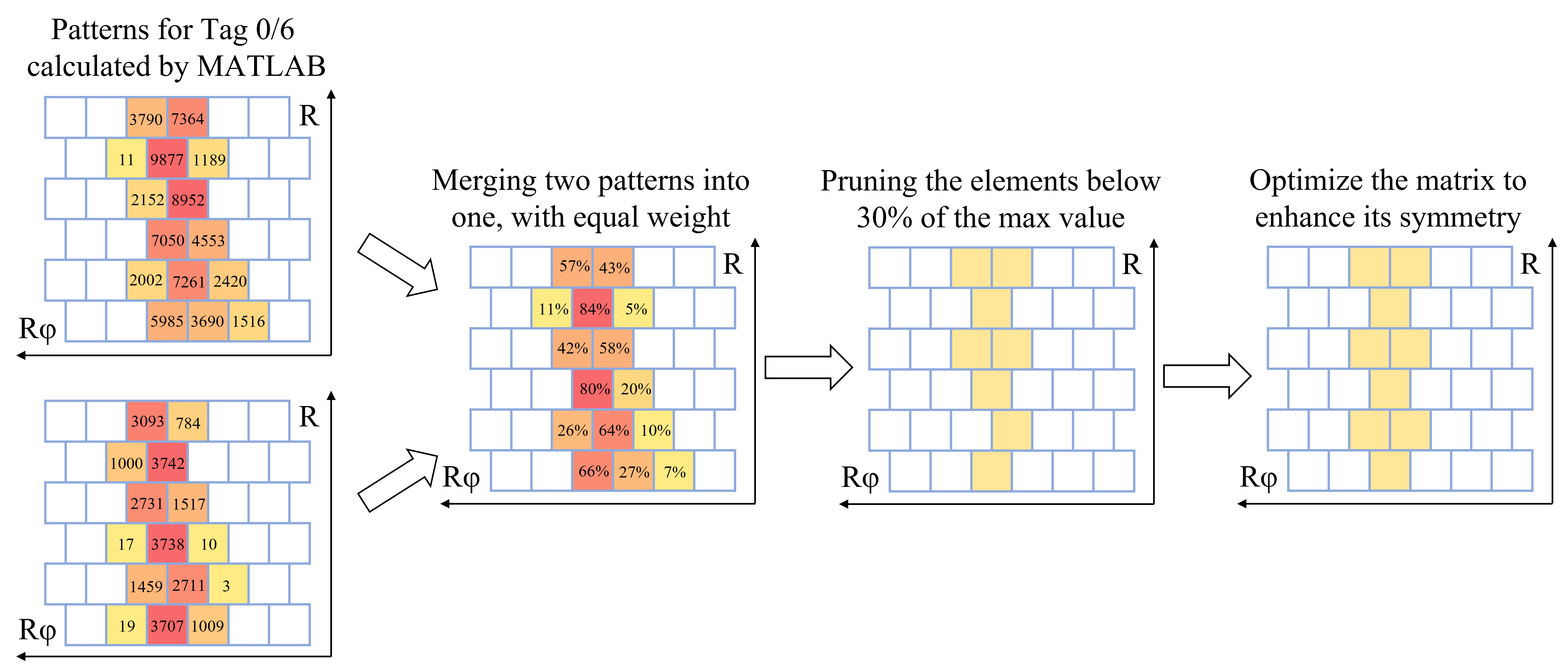} 
\caption{The schematic of patterns degeneracy step.} 
\label{fig:4}
\end{figure}

\begin{figure}[!t]
\centering
\includegraphics[width=\linewidth]{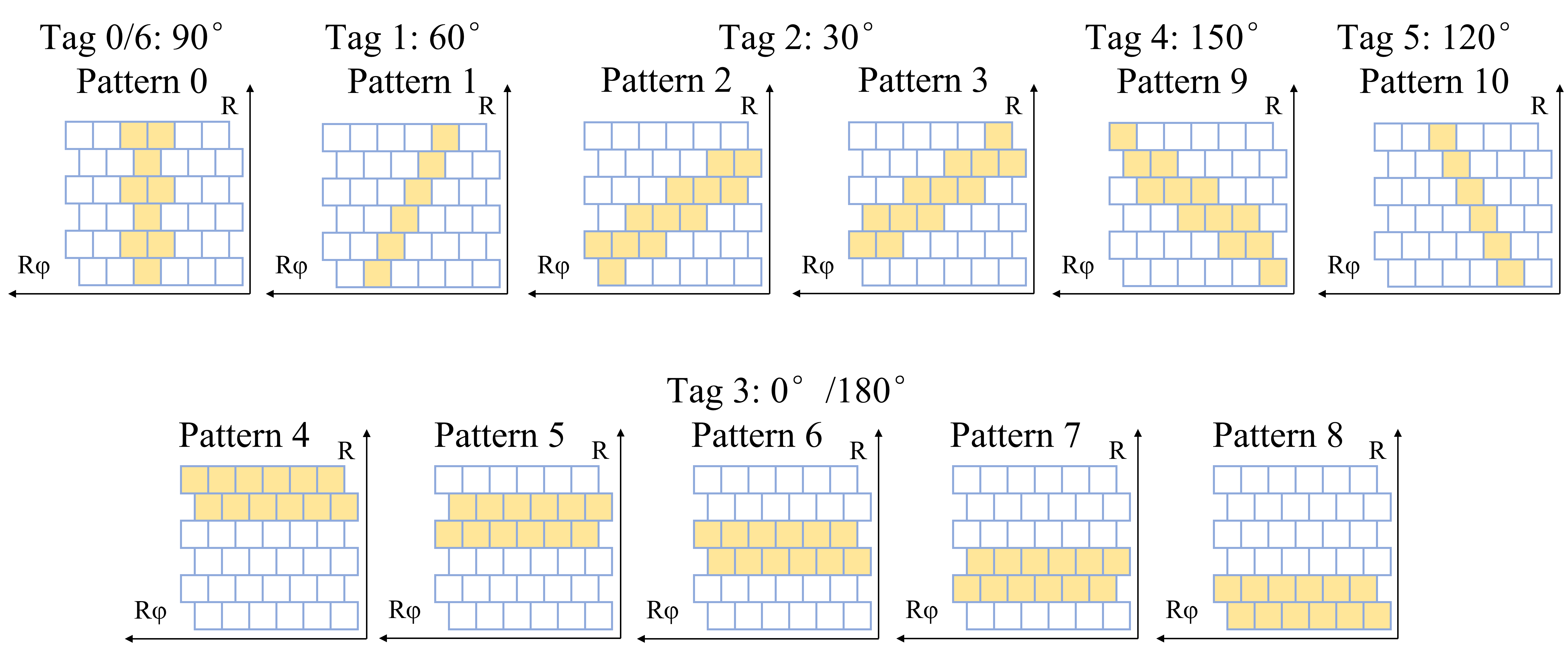} 
\caption{The determined 11 patterns with various tags.} 
\label{fig:5}
\end{figure}

The potential track segment shapes are calculated using MATLAB 2021a \cite{cite16} for the track with transverse momentum ranging from 50 to 2500 $MeV$, considering the geometry parameters of all superlayers. In this simulation, a candidate segment in a superlayer is encoded as a $6\times6$ matrix, representing the hit pattern across its six cell layers. By constraining the maximum number of hits in this matrix to 15, all geometrically feasible segment configurations can be classified into 53 distinct pattern templates. Each cell within the matrix is assigned a weight corresponding to its expected hit probability. Since the detection efficiency of the MDC cell is below 100\%, pattern cells with very low statistical weights can be pruned. This allows for further compression of the pattern bank, as illustrated in Fig.~\ref{fig:4}. The optimized pattern bank comprises 11 pattern templates, which cover all viable track segment shapes for transverse momentum ranging from 50 to 2500 MeV, as shown in Fig.~\ref{fig:5}.

In the track segment finding process, patterns with different tags are assigned distinct hit thresholds and weights, as listed in Table~\ref{tab:1}. Since the detection efficiency of the MDC cell is not 100\% in the experiment, the patterns must be capable of identifying track segments with a few missing hits. Therefore, the hit threshold is set to 4 or 5 hits for different patterns. Verification tests under different detection efficiencies are described in Section 5.3. Each pattern is assigned a different weight, while ensuring that a fully penetrating track segment receives a total weight of approximately 65 across all patterns. This weighting scheme enables consistent and comparable evaluation of different candidate paths during global tracking.

\begin{table}[!t]
\renewcommand{\arraystretch}{1.3}

\setlength{\tabcolsep}{5pt}

\caption{THE PARAMETERS FOR SEGMENT FINDING WITH EACH TAGS}
\label{tab:1}
\centering

\footnotesize
\begin{tabular}{c C{1cm} C{2cm} C{1cm} C{2cm}}
\toprule
Tag & \makecell{Pattern\\ID}  & \makecell{Hit\\threshold} & \makecell{Weight\\per hit} & \makecell{Additional\\requirement} \\
\midrule 

0/6 & 0 & 4 & 11 & Top \& bottom 3 layers each: $\ge$1 hit \\
1 & 1 & 4 & 17 & / \\
2 & 2 - 3 & 5 & 8 & Top \& bottom 3 layers each: $\ge$1 hit \\
3 & 4 - 8 & 5 & 10 & / \\
4 & 9 & 5 & 8 & Top \& bottom 3 layers each: $\ge$1 hit \\
5 & 10 & 4 & 17 & / \\

\bottomrule 
\end{tabular}
\end{table}

\begin{table}[!t]
\renewcommand{\arraystretch}{1.3}

\setlength{\tabcolsep}{3pt}

\caption{THE PATTERN BANK COVERAGE RATIO FOR SINGLE PARTICLES AND PHYSICS EVENTS}
\label{tab:2}
\centering

\footnotesize
\begin{tabular}{C{3cm} c c C{2cm}}
\toprule 
 \makecell{Single particle/physics\\events}  & $p_t(MeV/c)$ & $\theta(^\circ)$ & \makecell{Pattern bank\\coverage ratio} \\
\midrule 

    \multirow{5}{*}{$\mu^-$} & 200 - 2000 & 90 & 100\% \\
                                 & 50 - 100 & 90 & 99.5\% \\
                                 & 86.6 - 606 & 60 & 99.7\% \\
                                 & 50 - 350 & 30 & 96.8\% \\
                                 & 34.2 - 239.4 & 20 & 95.5\% \\
    \multirow{5}{*}{$\mu^+$} & 200 - 2000 & 90 & 100\% \\
                                 & 50 - 100 & 90 & 99.5\% \\
                                 & 86.6 - 606 & 60 & 99.7\% \\
                                 & 50 - 350 & 30 & 98.4\% \\
                                 & 34.2 - 239.4 & 20 & 96.8\% \\
    $K_S^0 \to \pi^+ \pi^-$ & 300 - 1500 & 20 - 160 & 99.1\% \\
    $J/\psi \to \Lambda \bar{\Lambda}, \Lambda \to p^+ \pi^-$ & / & / & 99.2\% \\
     \makecell{$J/\psi \to \Xi^+ \Xi^-,$\\$\Xi^- \to \Lambda \pi^-, \Lambda \to p \pi^-$} & / & / & 99.3\% \\

\bottomrule 
\end{tabular}
\end{table}

\begin{figure}[!t]
\centering
\includegraphics[width=0.8\linewidth]{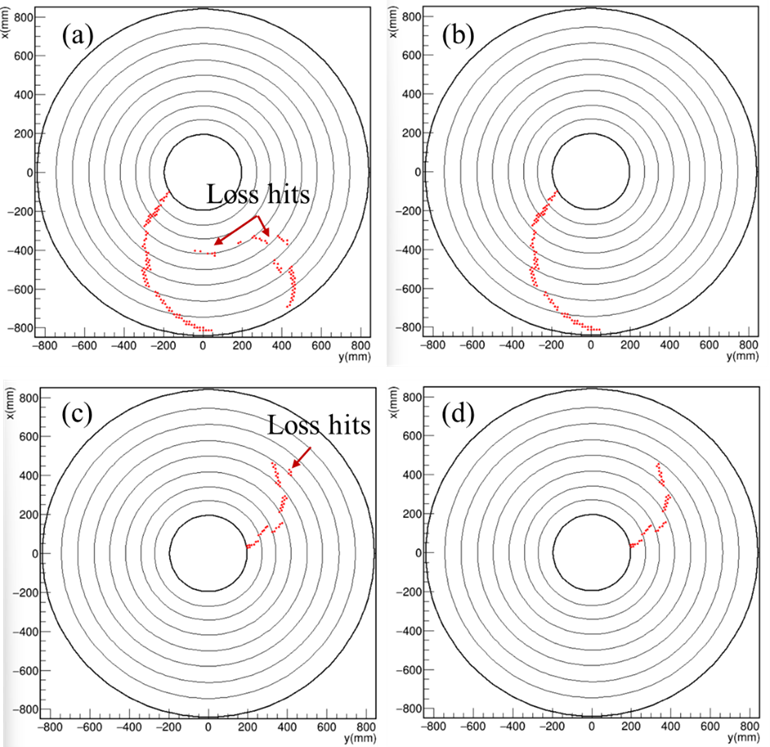} 
\caption{Typical cases of hit loss: (a–b) a looped track before and after pattern matching; (c–d) track with half-segment within a superlayer before and after pattern matching.} 
\label{fig:6} 
\end{figure}

To verify the effectiveness of the pattern bank, simulations are performed on the STCF software platform, Oscar \cite{cite17}. The simulations evaluate single-particle tracks across a wide range of transverse momentum and polar angles. In addition, pure signal events (without background) are generated to study displaced tracks, such as those from $J/\psi \to \Lambda \bar{\Lambda}, \Lambda \to p^+ \pi^-$ and $J/\psi \to \Xi^+ \Xi^-,\Xi^- \to \Lambda \pi^-, \Lambda \to p \pi^-$ processes. In these events, the decay vertex can be displaced significantly from the IP, with the largest separation reaching up to tens of cm. Table~\ref{tab:2} presents the simulated pattern bank coverage ratio for each case. This result indicates that the pattern bank can successfully identify all hits for the majority of charged tracks. As illustrated in Fig.~\ref{fig:6}, the loss of signal hits has two primary origins: the second half-turn of looped tracks, and half-segments within a superlayer. A cross-check with the STCF offline reconstruction algorithm \cite{cite18} confirms that these low-quality hits would be discarded. Consequently, the pattern bank selection introduces a negligible impact on offline tracking performance. Based on this analysis, the current pattern bank is validated for use in subsequent tracking stages.

\subsection{Edge matrix construction}

The segment connections in this algorithm are recorded within an edge matrix. A connection between two candidate segments, Segment 1 ($N_{SL1},Tag_1,\varphi_1,t_1$) and Segment 2 ($N_{SL2},Tag_2,\varphi_2,t_2$), is considered valid only if all of the following four criteria are satisfied:
\begin{enumerate}[label=\arabic*)]
    \item 	Layer proximity criterion: $(N_{SL2} - N_{SL1} = 1) \lor (N_{SL2} = N_{SL1} \land Tag_2 = 3)$

    This criterion enforces radial continuity. It requires segments to be in consecutive superlayers, ensuring a monotonic progression along the track radius. An exception is made for low-momentum tracks that may undergo a near-180° deflection; these are allowed to connect to another segment within the same superlayer, provided the downstream segment is labeled with Tag 3.
    
    \item 	Tag compatibility criterion: $\lvert Tag\_1 - Tag\_2 \rvert \leq 1$

    This limit ensures the transverse momentum changes continuously along a genuine track, which is reflected in the continuity of the associated pattern tags between adjacent segments.
    
    \item 	Azimuthal consistency criterion: $\varphi_{lower} \leq \varphi_1 - \varphi_2 \leq \varphi_{upper}$

    This positional threshold is derived from the allowable range of the azimuthal angle difference $\Delta\varphi$ between two segments. The thresholds $\varphi_{lower}$ and $\varphi_{upper}$ incorporate contributions from the stereo superlayer twist angle ($\Delta\varphi_{SL\_pre}$), magnetic-field-induced track deflection ($\Delta\varphi_{SLTag\_min}$ and $\Delta\varphi_{SLTag\_max}$), and segment measurement uncertainties ($MU$, $\Delta\varphi_{MU}$), as shown in \eqref{eq:1}.

    \item 	Temporal coincidence criterion: $T_0 - 50 \leq t_1, t_2 \leq T_0 + 550 \text{ (units: ns)}$

    This applies a common time window to select temporally coincident segments. The chosen window of $[-50, +550]$ ns relative to the event start time $T_0$ accommodates both the timing resolution and the maximum electron drift time within the MDC cells.
\end{enumerate}

\begin{equation}
\begin{aligned}
\varphi_{lower} &= - \Delta\varphi_{SL\_pre} + \Delta\varphi_{SLTag\_{min\_pre}} - \Delta\varphi_{MU\_pre}  \\
&\quad + \Delta\varphi_{SLTag\_{min\_post}} - \Delta\varphi_{MU\_post} \\
\varphi_{upper} &= \Delta\varphi_{SL_{pre}} + \Delta\varphi_{SLTag\_{max\_pre}} + \Delta\varphi_{MU\_pre} \\
&\quad + \Delta\varphi_{SLTag\_{max\_post}}  + \Delta\varphi_{MU\_post},
\end{aligned}
\label{eq:1}
\end{equation}

\begin{figure}[!t]
\centering
\includegraphics[width=\linewidth]{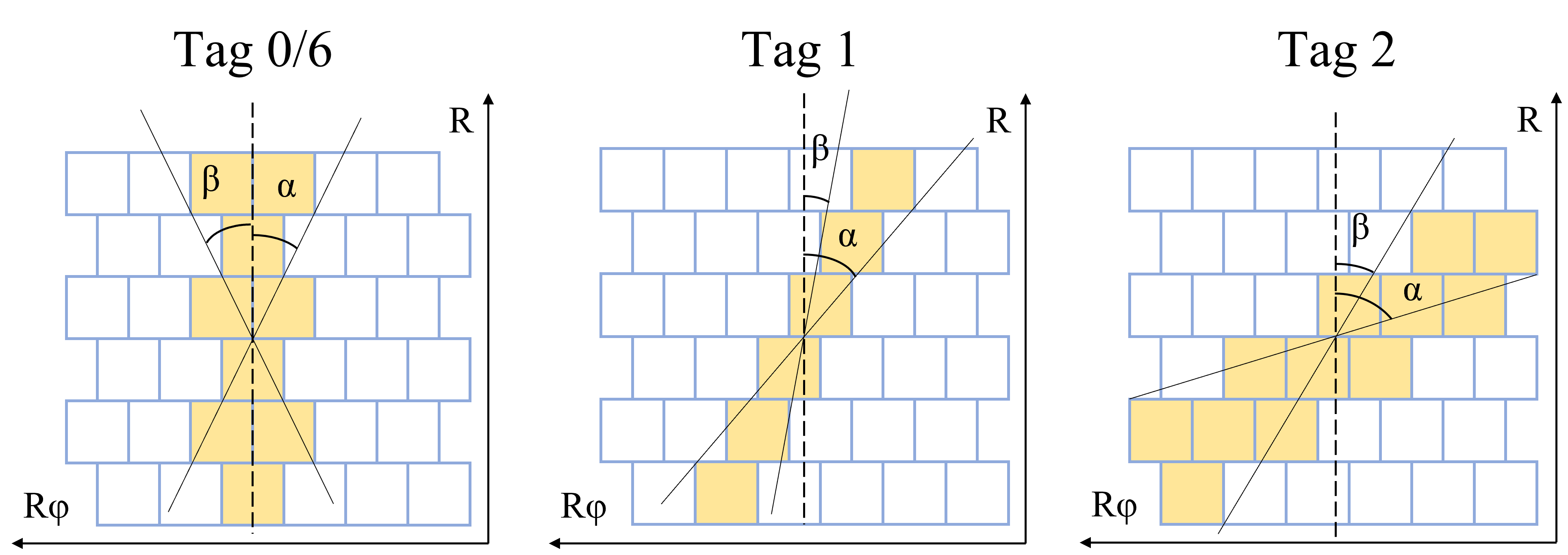} 
\caption{Maximum and minimum angles between segment tangent and radial line for different tags.} 
\label{fig:7} 
\end{figure}

Table~\ref{tab:3} summarizes the superlayer parameters for the STCF MDC conceptual design \cite{cite10}, and the threshold setting in edge matrix calculation. Based on these parameters, the threshold for the azimuth angle residual between adjacent superlayers ($\Delta\varphi_{SL\_pre}$) can be determined by accounting for the maximum twist angle introduced by the stereo superlayers in the X–Y plane. Additionally, the measurement uncertainty ($\Delta\varphi_{MU}$) can be estimated by assuming an error equivalent to one drift cell per superlayer. The azimuth angle changing from track deflection is related to the superlayer radius and the transverse momentum of the track. For each pattern, the angular range between the segment tangent and the radial line can be determined, as shown in Fig.~\ref{fig:7} and Table~\ref{tab:4}. Using these results, the maximum and minimum azimuth angle differences between adjacent track segments under all possible conditions are derived as \eqref{eq:2}, where $\alpha$ and $\beta$ denote the angular range of the specific tag, $R_{in}$ and $R_{out}$ are the radii of the superlayer, respectively.

\begin{table}[htbp]
  \renewcommand{\arraystretch}{1.3}

  \setlength{\tabcolsep}{0pt}
  \centering
  \caption{THE GEOMETRIC PARAMETERS OF MDC AND THE CORRESPONDING $\Delta \varphi$ FOR EACH SUPERLAYER}
  \label{tab:3}
  \begin{tabular}{c C{1cm} C{1cm} C{1.2cm} C{1cm} C{1.5cm} C{1.4cm} C{1.4cm}}
    \toprule
    SL & \makecell{$R_{in}$\\(mm)} & \makecell{$R_{out}$\\(mm)} & \makecell{Stereo\\angle\\($^\circ$)} & $N_{cell}$ & \makecell{Max. $\Delta\varphi$\\in $Z_{max}$\\($^\circ$)} & \makecell{$\Delta\varphi_{SL_{pre}}$\\($^\circ$)} & \makecell{$\Delta\varphi_{MU}$\\($^\circ$)} \\
    \midrule
    1 & 200.0 & 271.6 & 0 & 128 & 0 & 8 & 2.81 \\
    2 & 271.6 & 342.2 & \makecell{2.25\\to 2.73} & 160 & \makecell{6.15\\to 7.45} & 16 & 2.25 \\
    3 & 342.2 & 419.2 & \makecell{-2.36\\to -2.77} & 192 & \makecell{-6.4\\to -7.54} & 8 & 1.88 \\
    4 & 419.2 & 499.8 & 0 & 224 & 0 & 7 & 1.61 \\
    5 & 499.8 & 578.1 & \makecell{2.86\\to 3.23} & 256 & \makecell{6.63\\to 6.72} & 14 & 1.41 \\
    6 & 578.1 & 662.0 & \makecell{-2.94\\to 3.28} & 288 & \makecell{-5.98\\to -6.06} & 7 & 1.25 \\
    7 & 662.0 & 744.0 & 0 & 320 & 0 & 0 & 1.13 \\
    8 & 744.0 & 827.3 & 0 & 356 & 0 & / & 1.02 \\
    \bottomrule
  \end{tabular}
\end{table}

\begin{table}[htbp]
    \renewcommand{\arraystretch}{1.3}
    \setlength{\tabcolsep}{10pt}
    \centering
    \caption{THE CALCULATED MAXIMUM AND MINIMUM ANGLES FOR DIFFERENT TAGS}
    \label{tab:4}
    \begin{tabular}{lcccccc}
        \toprule
        Tag & 0/6 & 1 & 2 & 3 & 4 & 5 \\
        \midrule 
        $\alpha(^\circ)$ & -25 & -40 & -72 & $\pm 68$ & 72 & 40 \\
        $\beta(^\circ)$ & 25 & -10 & -35 & $\pm 112$ & 35 & 10 \\
        \bottomrule 
    \end{tabular}
\end{table}

\begin{equation}
\begin{aligned}
\Delta\varphi_{SLTag\_min} &= \tan(\beta) \cdot \frac{R_{out} - R_{in}}{2 \cdot R_{out}} \\
\Delta\varphi_{SLTag\_max} &= \tan(\alpha) \cdot \frac{R_{out} - R_{in}}{2 \cdot R_{out}},
\end{aligned}
\label{eq:2}
\end{equation}

With these criteria, the edge matrix is constructed as a sparse matrix. The density of this matrix is related to the occupancy of the MDC. Taking the process $e^+ e^- \to \pi^+ \pi^- J/\psi,  J/\psi \to e^+ e^-$ as an example, Table~\ref{tab:5} presents the number of identified track segments, the count of valid edges, and the resulting density of the edge matrix. In the simulation, signal hits are mixed with background at various safety factors, where the background contributions include both luminosity-related processes (Bhabha scattering, di-gamma, and di-$\mu$ events) and beam-related effects (Touschek effect, column scattering, and bremsstrahlung) \cite{cite19}. The simulated data cover a time window of 800 ns, matching the sampling window width of the STCF MDC. This result demonstrates that even at occupancies exceeding 15\%, this segment-based HLT tracking algorithm can still produce a highly sparse edge matrix, thereby significantly reducing the complexity of global tracking.

\begin{table}[htbp]
  \renewcommand{\arraystretch}{1.3}

  \setlength{\tabcolsep}{3pt}
  \centering
  \caption{THE PERFORMANCE OF SEGMENT FINDING AND EDGE CONSTRUCTION WITH $e^+ e^- \to \pi^+ \pi^- J/\psi,  J/\psi \to e^+ e^-$ UNDER DIFFERENT OCCUPANCY}
  \label{tab:5}
  \begin{tabular}{C{1.5cm} C{1.5cm} C{1.5cm} C{1cm} C{2cm}}
    \toprule
    \makecell{Safety\\factor} & \makecell{MDC\\occupancy} & \makecell{No. of\\identified\\segments} & \makecell{No. of\\valid\\edges} & \makecell{Density of\\edge matrix} \\
    \midrule
    $\times$ 1 bkg & 4.1\% & 60.6 & 57.5 & 1.57\% \\
    $\times$ 3 bkg & 8.3\% & 107.8 & 122.4 & 1.05\% \\
    $\times$ 5 bkg & 12.2\% & 163.9 & 229.9 & 0.86\% \\
    $\times$ 8 bkg & 18.8\% & 260.5 & 498.4 & 0.73\% \\
    $\times$ 10 bkg & 22.4\% & 327.1 & 718.3 & 0.67\% \\
    $\times$ 12 bkg & 25.7\% & 393.2 & 942 & 0.61\% \\
    \bottomrule
  \end{tabular}
\end{table}

\subsection{Stereo superlayer segment pair merging}

\begin{figure}[!t]
\centering
\includegraphics[width=\linewidth]{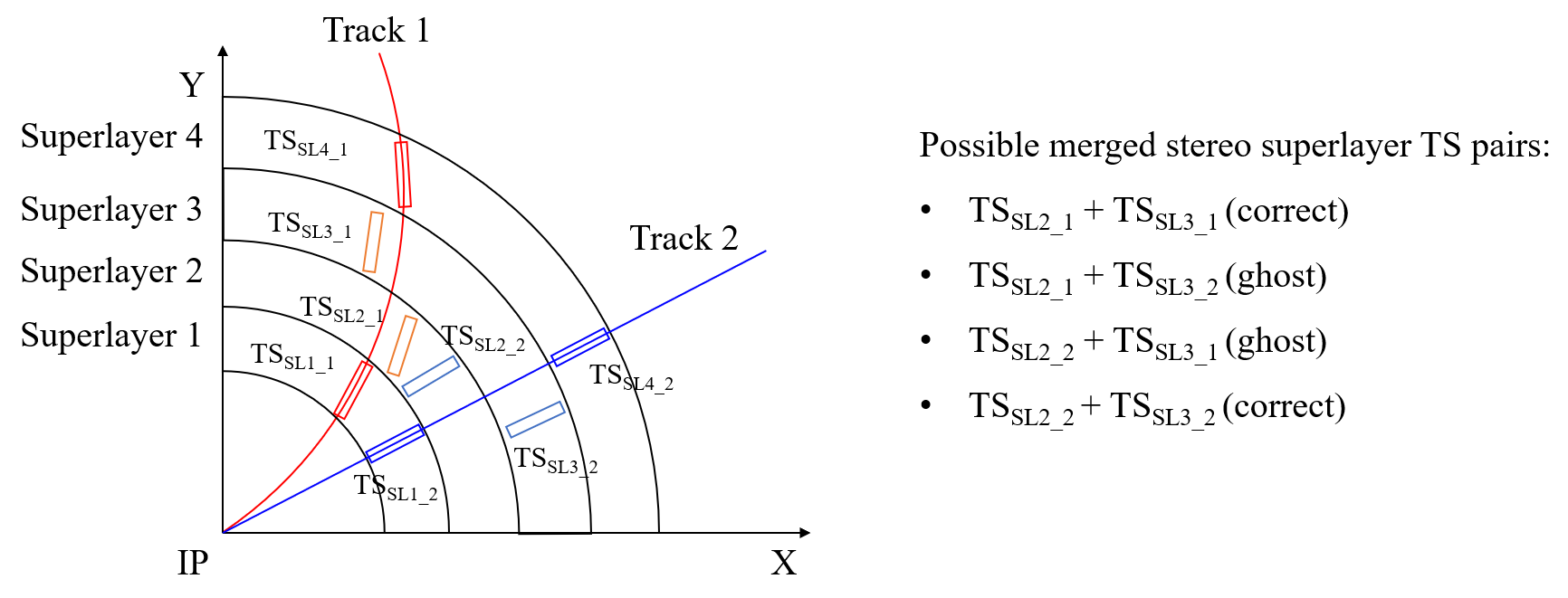} 
\caption{Conceptual diagram of stereo superlayer segments merging.}
\label{fig:8} 
\end{figure}

The twisted angle of stereo superlayers enables the measurement and analysis of a track's polar angle. However, this configuration also leads to discontinuities in the azimuth angle of track segments, which interferes with straightforward circular track finding. To facilitate the calculation of track shape and verification of tracking quality, such as through Hough transform or least squares method (LSM), pairs of connected segments in stereo superlayers are merged into a single segment. As illustrated in Fig.~\ref{fig:8}, this merging process not only generates correct stereo superlayer segment pairs but also introduces some “ghost segments”. In the subsequent global tracking stage, all candidate tracks are evaluated using LSM, and ghost segments are filtered out as they fail to pass the residual threshold, exerting minimal influence on the final tracking result. Figure~9 shows the relationship between detector occupancy, the number of track segments, the count of valid edges, and the density before and after merging. At low occupancy, the merging of stereo superlayer segments reduces the number of both segments and valid edges. Even at high occupancy, although the edge count increases moderately, the density of the edge matrix remains below 1\%. This demonstrates that the stereo superlayer segment merging maintains a highly sparse data structure and is therefore fully acceptable under a wide range of occupancy conditions.

\begin{figure}[!t] 
    \centering
    
    \begin{subfigure}[b]{\linewidth} 
        \centering
        \includegraphics[width=0.8\linewidth]{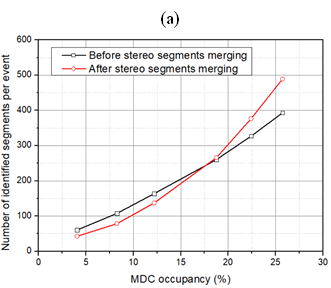}
    \end{subfigure}
    
    \vspace{0cm} 
    
    \begin{subfigure}[b]{\linewidth}
        \centering
        \includegraphics[width=0.8\linewidth]{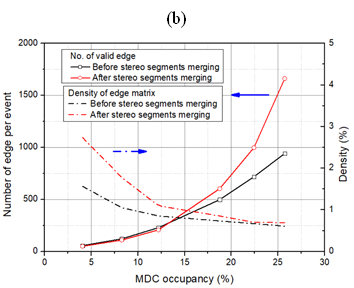}
    \end{subfigure}
    
    \caption{The number of segments, valid edges, and density of edge matrix before and after segments merging.}
    \label{fig:9}
\end{figure}

After segment merging, the key parameters of each resulting segment are compiled into a structured list to support efficient global track reconstruction, including the superlayer number, azimuth angle, pattern tag, hit count within the pattern, and weight.

\section{GLOBAL TRACKING}

\begin{figure}[!t]
\centering
\includegraphics[width=\linewidth]{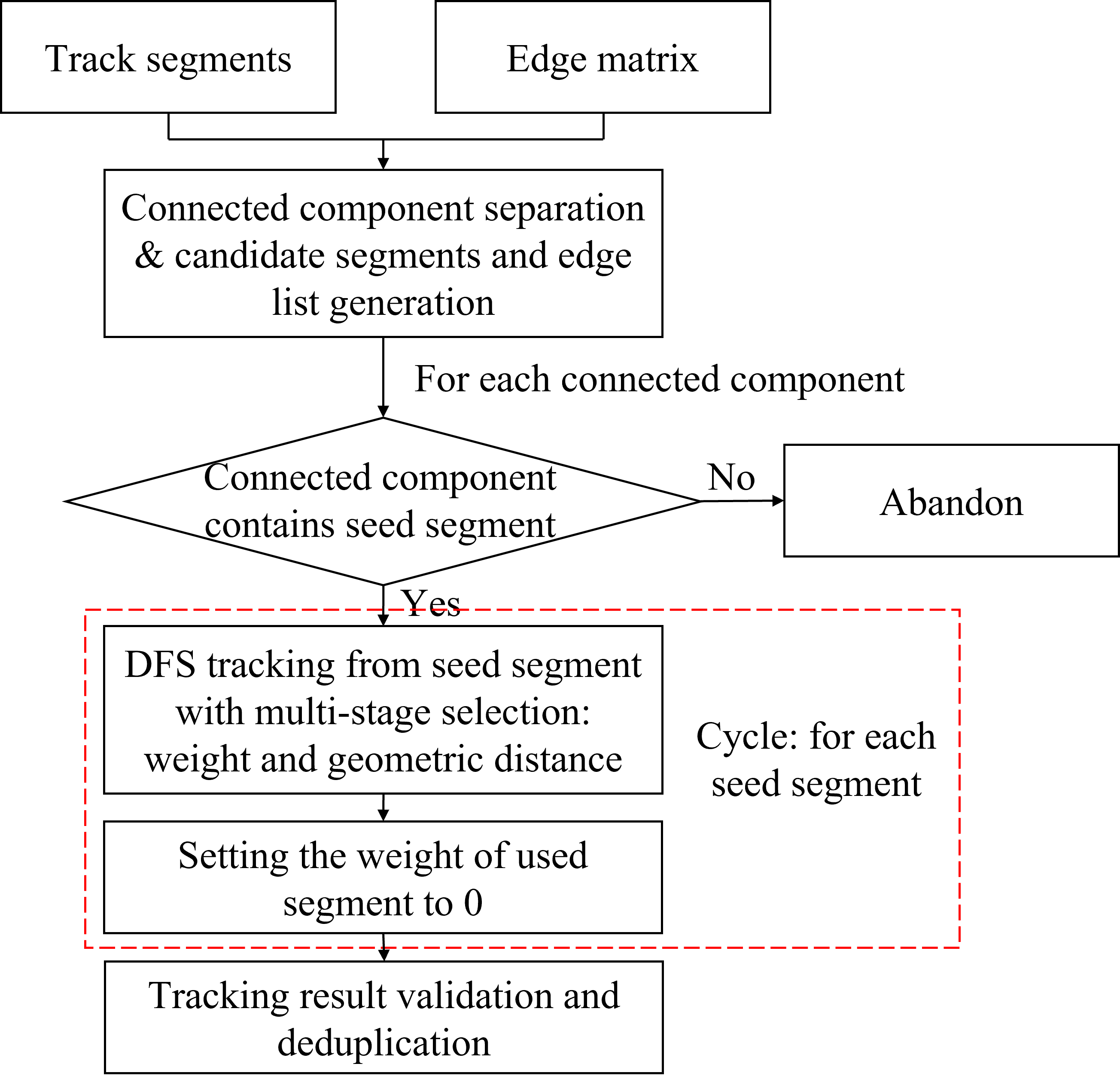} 
\caption{ Flowchart of the global tracking based on DFS.}
\label{fig:10}
\end{figure}

Once all track segments are identified and their connections have been recorded in an edge matrix, the algorithm proceeds to the global tracking stage. A variety of technical approaches can be applied for global track reconstruction, including connected component analysis using DFS and graph neural networks (GNNs). In this study, global tracking is implemented by performing DFS within each connected component, as illustrated in Fig.~\ref{fig:10}.

\begin{figure}[!t] 
    \centering

    \begin{subfigure}[b]{\linewidth} 
        \centering
        \includegraphics[width=0.6\linewidth]{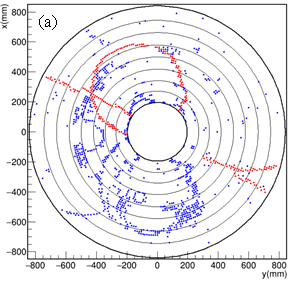} 
    \end{subfigure}
    
    \vspace{0cm} 
    
    \begin{subfigure}[b]{\linewidth}
        \centering
        \includegraphics[width=0.6\linewidth]{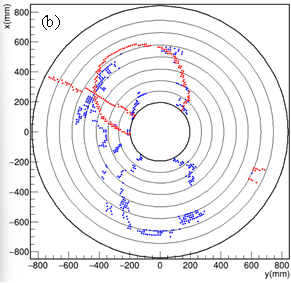}
    \end{subfigure}

    \vspace{0cm} 

    \begin{subfigure}[b]{\linewidth}
        \centering
        \includegraphics[width=0.6\linewidth]{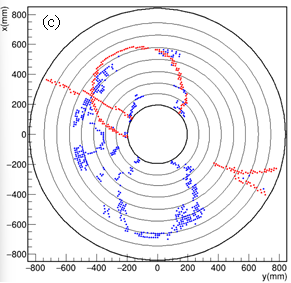}
    \end{subfigure}
    
    \caption{A typical $J/\psi \to \Lambda \bar{\Lambda}, \Lambda \to p^+ \pi^-$ event. Signal hits are marked in red and background hits in blue: (a) Raw hits in the MDC. (b) HLT tracking result without using seed segments from superlayer 4. (c) HLT tracking result when including seed segments from superlayer 4.}
    \label{fig:11}
\end{figure}

All the edges are grouped into several connected components, with their corresponding segments recorded in the candidate lists. Within each connected component, the seed segments are selected from superlayer 1 and superlayer 4. For standard prompt tracks, seed segments from superlayer 1 are sufficient, whereas some of the long-lived particles require seeds from superlayer 4, as illustrated in Fig.~\ref{fig:11}. Any connected component that contains no valid seed segment is discarded. To handle potential track crossing in superlayer 1 or 4, a seed segment is duplicated if its hit count exceeds 15. During tracking process, each seed segment initiates three independent tracking attempts, corresponding to three pre-defined track hypotheses: a high‑$p_t$ track, a low‑$p_t$ track with positive charge, and a low‑$p_t$ track with negative charge. After DFS-based tracking, duplicated and low‑quality candidate tracks within the same connected component are eliminated through a deduplication process.

\subsection{DFS tracking}

There is a trade-off between tracking efficiency and data compression ratio in DFS-based tracking. Thus, the optimal path for each seed segment is determined through a multi-stage selection process.

First, three independent tracking attempts are performed under distinct hypotheses: a high-$p_t$ track, a low-$p_t$ track with positive charge, and a low-$p_t$ track with negative charge, as shown in Fig.~12. This design allows a single seed segment to yield up to three candidate tracks, thereby comprehensively covering potential extreme crossing track cases. For the high-$p_t$ track hypothesis, every segment along the candidate path must carry Tag 0. For the low-$p_t$ positive‑charge hypothesis, at least one segment in the path is labeled with Tag 1. Similarly, the low-$p_t$ negative‑charge hypothesis requires at least one segment labeled with Tag 5. This approach controls the size of connected components, thereby improving both the signal retention ratio and the computational speed under high occupancy. 

\begin{figure}[!t]
\centering
\includegraphics[width=\linewidth]{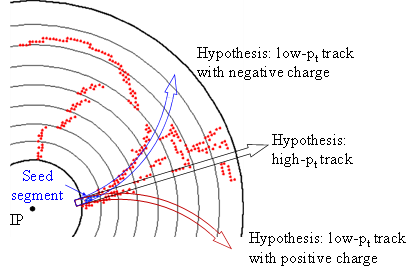} 
\caption{The schematic of the three hypotheses in tracking.} 
\label{fig:12}
\end{figure}

Second, during the DFS for path segments, if a candidate path penetrates beyond superlayer 6, the segments from the two stereo superlayer pairs are cross‑checked for geometric consistency. As shown in~\ref{eq:3}, the azimuth angle of each stereo superlayer is predicted under the hypothesis that the polar angle equals $90^\circ$, using the superlayer indices available in the candidate path. In this equation, $\varphi_i$ denotes the measured azimuth angle of the segment in superlayer $i$, and $\varphi_{i\_0}$ is the predicted azimuth angle for the stereo superlayers under the $90^\circ$‑polar‑angle hypothesis. For a genuine track with an unchanged polar angle, the difference between $\varphi_{i\_0}$ and $\varphi_i$ is determined by the geometry parameters of the four stereo superlayers, as shown in \eqref{eq:4}. In this equation, $N_i$ is the number of cells in superlayer $i$, $\Delta N_i$ is the maximum twist (in cells) of that superlayer, $R_i$ is its radius, and $C_i$ represents a constant value for the path. All the geometry parameters can be found in Table~\ref{tab:3}. This relationship forms one of the key rationales for the stereo superlayer design in the drift chamber. Therefore, the geometric consistency check can be performed by computing the ratio $(C_6 \cdot C_2) / (C_5 \cdot C_3)$ for the candidate path. If the resulting value lies within the interval $[0.8,1.2]$, the path is accepted.

\begin{equation}
\begin{aligned}
    \varphi_{2\_0} &= \frac{\varphi_2 + \varphi_3}{2} - \left( \varphi_4 - \frac{\varphi_2 + \varphi_3}{2} \right) \cdot 0.3 \\
    \varphi_{3\_0} &= \frac{\varphi_2 + \varphi_3}{2} + \left( \varphi_4 - \frac{\varphi_2 + \varphi_3}{2} \right) \cdot 0.37 \\
    \varphi_{5\_0} &= \frac{\varphi_5 + \varphi_6}{2} - \left( \varphi_7 - \frac{\varphi_5 + \varphi_6}{2} \right) \cdot 0.3 \\
    \varphi_{6\_0} &= \frac{\varphi_5 + \varphi_6}{2} - \left( \varphi_7 - \frac{\varphi_5 + \varphi_6}{2} \right) \cdot 0.37,
\end{aligned}
\label{eq:3}
\end{equation}

\begin{equation}
    \left| \varphi_{i\_0} - \varphi_i \right| \cdot \frac{N_i}{\Delta N_i \cdot R_i} = C_i,
    \label{eq:4}
\end{equation}

Third, candidate paths containing more than three segments are fitted either to a circular track or to a straight line using the LSM. The LSM determines the fitted track by minimizing the algebraic distance. If the radius of fitted circular track exceeds 500 mm, the straight-line fit is applied as a candidate, in order to accommodate the straight high‑$p_t$ tracks. Based on the fitted tracks, the mean square distance (MSD) is then evaluated according to \eqref{eq:5}, taking the minimum value obtained from the circular and straight‑line fits. In this equation, n denotes the number of segments in the path, $(x_i, y_i)$ are the coordinates of each segment, $(x_0, y_0)$ and $r_0$ represent the center and radius of the fitted circle, respectively, while $A$, $B$ and $C$ are the parameters of the fitted straight line, normalized such that $A^2 + B^2 = 1$.

\begin{equation}
\begin{aligned}
    MSD_{circular} &= \frac{1}{n} \cdot \sum \left( \sqrt{(x_i - x_o)^2 + (y_i - y_o)^2} - r_0 \right)^2 \\
    MSD_{line} &= \frac{1}{n} \cdot \sum (A \cdot x_i + B \cdot y_i + C)^2.
\end{aligned}
\label{eq:5}
\end{equation}

Segments with a tag value between 1.5 and 4.5 are classified as “curved segments”. If a candidate path contains more than two curved segments, the MSD threshold for rejection is raised to 150. Otherwise, the default threshold of 120 is applied. The optimal path is selected as follows: a candidate with a weight exceeding the current best by 20 is chosen immediately. Otherwise, a candidate with weight $>95\%$ of the current best and a smaller MSD replaces it. The total path weight is the sum of the weights of its constituent segments, and the MSD is computed using \eqref{eq:5}. Figure~13 presents the MSD distribution for single tracks within the transverse momentum range of 75–2000 $MeV/c$ and the polar angle range of $20^\circ–160^\circ$. The distribution confirms that the chosen rejection thresholds are appropriate. For the ultra low‑$p_t$ tracks (e.g., $50 MeV/c$), fewer than four segments are typically available. In such cases, the MSD metric is not applicable, and only weight criterion is applied.

\begin{figure}[!t]
\centering
\includegraphics[width=0.8\linewidth]{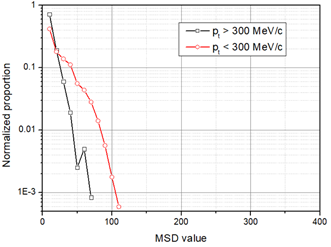} 
\caption{The MSD distribution for single tracks with different transverse momentum range.} 
\label{fig:13} 
\end{figure}

\subsection{Track deduplication}

All tracks reconstructed within a connected component are sorted in descending order of weight and then deduplicated, starting from the highest‑weight track. A track is discarded if it does not contribute at least one new axial‑superlayer segment to the overall hit assignment. Also, any track that lacks a matched stereo‑superlayer pair, such as an ultra‑low‑$p_t$ track penetrating only superlayers 1 and 2, is also rejected.

\section{PERFORMANCE EVALUATION}

The HLT tracking algorithm is designed for removing the background hits, compressing data storage, while maintain signal hits. Thus, the data compression ratio, signal retention ratio, data purity, and influence on offline tracking are the key performance parameters.

\subsection{Evaluation with pure signal tracks}

\begin{figure}[!t]
\centering
\includegraphics[width=\linewidth]{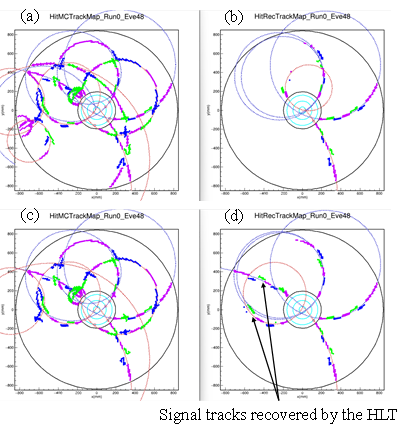} 
\caption{The offline reconstruction result without (a-b) and with (c-d) HLT tracking algorithm for a $J/\psi \to \Xi^- \Xi^+,\Xi^- \rightarrow \Lambda \pi^-,\Lambda \rightarrow p^+ \pi^-$ event. (a) Raw MDC input hits and truth signal tracks. (b) Reconstruction result with raw MDC input. (c) MDC hits filtered by HLT tracking algorithm. (d) Reconstruction result with HLT filtered MDC data.} 
\label{fig:14}
\end{figure}

Table~\ref{tab:6} presents the signal retention ratio for different types of single tracks or physics events. Since no background is included in this simulation, the data compression ratio is equal to the signal retention ratio. The simulated track types cover normal tracks with varying transverse momentum and polar angle, looped tracks, low‑$p_t$ tracks, and displaced tracks. It can be observed that certain track categories, such as looped track and displaced track, do not achieve an extreme high signal retention ratio. This is because the HLT tracking algorithm retains only the first half‑turn of looped tracks and discards the half‑segments within a superlayer, as illustrated in Fig.~\ref{fig:6}. To validate this logic, the standard offline reconstruction algorithm of the STCF is used to compare reconstruction results with and without the HLT tracking process. A representative comparison is presented in Fig.~\ref{fig:14}. It confirms that retaining only the first half‑turn of looped tracks is sufficient for offline tracking, and that low‑quality hits can be safely removed. Notably, the HLT reconstruction can even achieve higher tracking efficiency in some cases, because it filters out such low‑quality hits.

\begin{table}[htbp]
\renewcommand{\arraystretch}{1.3}

\setlength{\tabcolsep}{1pt}
\centering
\caption{THE SIMULATED SIGNAL RETENTION RATIO FOR THE TYPICAL SINGLE TRACKS AND PHYSICS EVENTS WITHOUT BACKGROUND}
\label{tab:6}
\begin{tabular}{C{1.4cm} C{2.5cm} c C{1.5cm} C{1.3cm}}
\toprule
Type & \makecell{Single\\particle/physics\\events} & $p_t (MeV/c)$ & \makecell{$\theta$\\($^\circ$)} & \makecell{Signal\\retention\\ratio} \\
\midrule

\multirow{10}{*}{\makecell{Normal\\track}}
    & \multirow{5}{*}{$\mu^-$} & 200 - 2000 & 90 & 99.9\% \\
    & & 50 - 100 & 90 & 92.2\% \\
    & & 86.6 - 606 & 60 & 97.7\% \\
    & & 50 - 350 & 30 & 96.0\% \\
    & & 34.2 - 239.4 & 20 & 94.2\% \\
    & \multirow{5}{*}{$\mu^+$} & 200 - 2000 & 90 & 99.9\% \\
    & & 50 - 100 & 90 & 91.1\% \\
    & & 86.6 - 606 & 60 & 96.7\% \\
    & & 50 - 350 & 30 & 96.6\% \\
    & & 34.2 - 239.4 & 20 & 95.7\% \\

\multirow{2}{*}{\makecell{Looped\\track}}
    & $\mu^-$ & 50 - 100 & 90 & 92.2\% \\
    & $\mu^+$ & 50 - 100 & 60 & 91.1\% \\

\multirow{2}{*}{\makecell{Low-$p_t$\\track}}
    & $\mu^-$ & 50 - 100 & 30 & 95.4\% \\
    & $\mu^+$ & 50 - 100 & 30 & 95.3\% \\

\multirow{5}{*}{\centering \makecell{Displaced\\track}}
    & $K_S^0 \to \pi^+ \pi^-$ & 300 - 1500 & 20 - 160 & 88.6\% \\
    & \makecell{$J/\psi \to \Lambda \bar{\Lambda},$\\$\Lambda \rightarrow p^+ \pi^- $} & / & / & 99.9\% \\
    & \makecell{$J/\psi \to \Xi^- \Xi^+,$\\$\Xi^- \rightarrow \Lambda \pi^-,$\\$\Lambda \rightarrow p^+ \pi^-$} & / & / & 97.7\% \\ 
    
\bottomrule
\end{tabular}
\end{table}

\subsection{Physics events with background evaluation}

\begin{figure}[!t] 
    \centering

    \begin{subfigure}[b]{\linewidth} 
        \centering
        \includegraphics[width=0.6\linewidth]{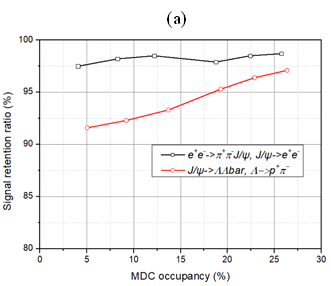} 
    \end{subfigure}
    
    \vspace{0cm} 

    \begin{subfigure}[b]{\linewidth}
        \centering
        \includegraphics[width=0.6\linewidth]{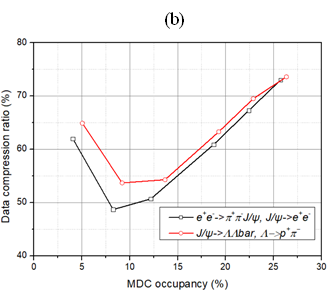}
    \end{subfigure}
    
    \caption{The signal retention ratio and data compression ratio for the physics events, with various MDC occupancy.}
    \label{fig:15}
\end{figure}

To evaluate the performance of the proposed algorithm, the physics reaction of $e^+ e^-\rightarrow \pi^+ \pi^- J/\psi,J/\psi \rightarrow e^+ e^-$ and $J/\psi \to \Lambda \bar{\Lambda},\Lambda \rightarrow p^+ \pi^- $ are tested at different detector occupancies by varying the background safety factor from 1 to 12. The $e^+ e^- \rightarrow \pi^+ \pi^- J/\psi,J/\psi \rightarrow e^+ e^-$ is a typical physical process commonly used to test the tracking performance of the tracker system, including cases such as high-$p_t$, low-$p_t$, and crossing tracks. The $J/\psi \to \Lambda \bar{\Lambda},\Lambda \rightarrow p^+ \pi^- $ provides displaced tracks with a range of decay lengths and looped tracks for the algorithm performance evaluation. Figure~\ref{fig:15} shows the signal retention ratio and data compression ratio under different MDC occupancy levels. This results demonstrate that the novel HLT tracking algorithm maintains a stable signal retention ratio even at high occupancies above 25\%. The signal retention ratio increases with occupancy, as more background segments contribute to track formation while low-quality signal hits are also recorded. The data compression ratio initially decreases and then rises with increasing occupancy. In the low-occupancy region (e.g., 5\%), most MDC hits are signal hits, while background hits can be efficiently rejected; the data compression ratio in this case is close to the signal purity of the input MDC data. In the high-occupancy region, a large fraction of background hits is retained to achieve a high signal retention ratio, leading to an increase in data compression ratio as the MDC occupancy rises.

\begin{figure}[!t]
\centering
\includegraphics[width=0.6\linewidth]{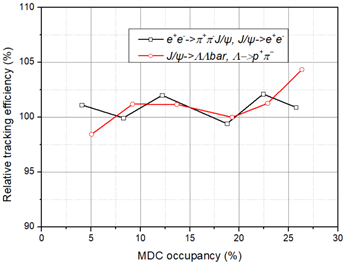} 
\caption{The relative tracking efficiency with proposed HLT tracking algorithm, evaluated by STCF offline reconstruction algorithm.} 
\label{fig:16} 
\end{figure}

\begin{figure}[!t] 
    \centering

    \begin{subfigure}[b]{\linewidth} 
        \centering
        \includegraphics[width=0.6\linewidth]{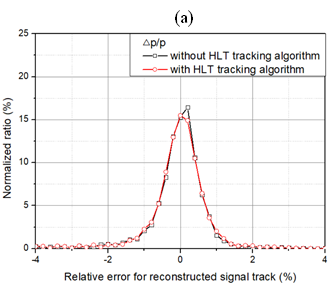}
    \end{subfigure}
    
    \vspace{0cm}

    \begin{subfigure}[b]{\linewidth}
        \centering
        \includegraphics[width=0.6\linewidth]{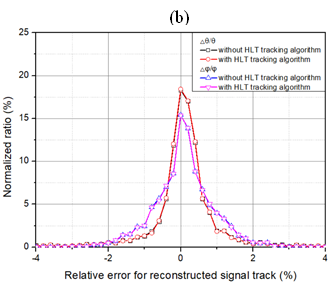}
    \end{subfigure}
    
    \caption{The relative error for reconstructed signal track in $e^+ e^-\rightarrow \pi^+ \pi^- J/\psi,J/\psi \rightarrow e^+ e^-$.}
    \label{fig:17}
\end{figure}

\begin{figure}[!t] 
    \centering

    \begin{subfigure}[b]{\linewidth} 
        \centering
        \includegraphics[width=0.6\linewidth]{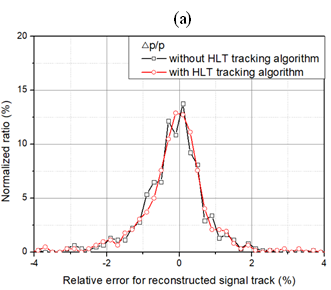} 
    \end{subfigure}
    
    \vspace{0cm}

    \begin{subfigure}[b]{\linewidth}
        \centering
        \includegraphics[width=0.6\linewidth]{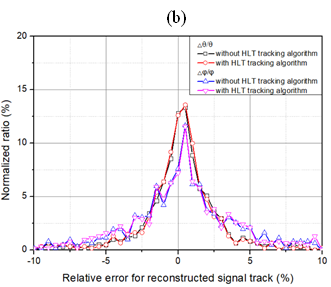}
    \end{subfigure}
    
    \caption{The relative error for reconstructed signal track in $J/\psi \to \Lambda \bar{\Lambda},\Lambda \rightarrow p^+ \pi^- $.}
    \label{fig:18}
\end{figure}

\begin{figure}[!t]
\centering
\includegraphics[width=\linewidth]{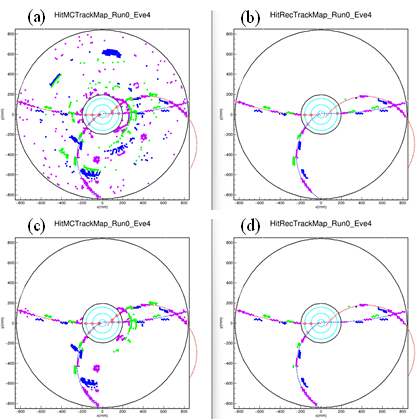} 
\caption{The offline reconstruction result without (a-b) and with (c-d) HLT tracking algorithm for a $J/\psi \to \Lambda \bar{\Lambda},\Lambda \rightarrow p^+ \pi^- $ event with 13\% occupancy. (a) Raw MDC input hits and truth signal tracks. (b) Reconstruction result with raw MDC input. (c) MDC hits filtered by HLT tracking algorithm. (d) Reconstruction result with HLT filtered MDC data.}
\label{fig:19}
\end{figure}

\begin{figure}[!t]
\centering
\includegraphics[width=\linewidth]{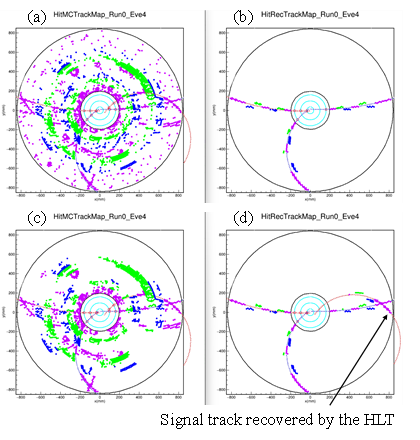} 
\caption{The offline reconstruction result without (a-b) and with (c-d) HLT tracking algorithm for a $J/\psi \to \Lambda \bar{\Lambda},\Lambda \rightarrow p^+ \pi^- $ event with 27\% occupancy. (a) Raw MDC input hits and truth signal tracks. (b) Reconstruction result with raw MDC input. (c) MDC hits filtered by HLT tracking algorithm. (d) Reconstruction result with HLT filtered MDC data.} 
\label{fig:20} 
\end{figure}

In the evaluation, the standard STCF offline reconstruction algorithm serves as the benchmark. The comparison focuses on the tracking efficiency to the signal tracks, as well as the relative error distributions of the reconstructed momentum, polar angle, and azimuth angle. When the tracking efficiency achieved by the offline reconstruction (without HLT tracking) is normalized to 100\%, the relative tracking efficiency with the HLT algorithm exhibits negligible difference, as shown in Fig.~\ref{fig:16}. Figure~\ref{fig:17} and Fig.~\ref{fig:18} also presents the relative error distributions of the reconstructed momentum, polar angle, and azimuth angle with the background safety factor of 3 (around 8\% occupancy), demonstrating that the reconstruction parameters remain consistent. Figure~\ref{fig:19} and Fig.~\ref{fig:20} present a typical $J/\psi \to \Lambda \bar{\Lambda},\Lambda \rightarrow p^+ \pi^- $ event with 13\% and 27\% occupancy, demonstrating the good and stable HLT tracking performance. These results confirm that the proposed HLT tracking algorithm does not introduce any significant adverse effects on the subsequent offline analysis.

\subsection{Performance verifying with various detection efficiency}

\begin{figure}[!t]
\centering
\includegraphics[width=0.6\linewidth]{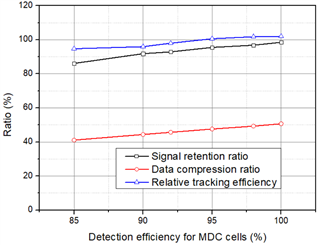} 
\caption{The performance of HLT tracking algorithm with various detection efficiency and approximately 12\% occupancy.} 
\label{fig:21}
\end{figure}

The detection efficiency of an MDC cell typically exceeds 98\% under normal operating conditions, but can drop to around 95\% or lower after extended operation and gas aging. Therefore, the HLT tracking algorithm must perform stably across a range of detection efficiencies. A physics event sample of $e^+ e^-\rightarrow \pi^+ \pi^- J/\psi,J/\psi \rightarrow e^+ e^-$ is simulated with the assumption of detection efficiencies ranging from 85\% to 100\%. In this simulation, the background safety factor is set to 5 to simulate more challenging working conditions, resulting in an occupancy of approximately 13\%. Figure~\ref{fig:21} presents the signal retention ratio, data compression ratio, and the relative tracking efficiency evaluated by the offline reconstruction algorithm under these conditions. It can be observed that when the detection efficiency exceeds 92\%, the signal retention ratio of the HLT tracking algorithm remains above 93\%, while the relative tracking efficiency reaches 98\%, demonstrating stable performance. This result indicates that the proposed HLT tracking algorithm performs reliably when the detection efficiency is above 92\%.

\section{DISCUSSION}

The proposed HLT tracking algorithm is designed based on track segment, which significantly reduces the element number and edge matrix size required for global tracking. The processes of track segment finding and edge matrix construction can be further accelerated by optimizing matrix operations and adopting a highly parallelized algorithm architecture. However, global tracking involves complex logical calculations, which constrain the overall computing latency, particularly under extremely high MDC occupancy. For instance, the computing latency is approximately 200 ms for a typical $e^+ e^-\rightarrow \pi^+ \pi^- J/\psi,J/\psi \rightarrow e^+ e^-$ event at around 8\% occupancy. To reduce latency under high-occupancy conditions, two main approaches are being considered. The first involves applying a neural network to identify the optimal path within connected components, though this is challenging due to the variable number of target tracks. The second method focuses on discarding more low-quality edges between segments, such as by using the Hough transform to derive more precise momentum angles, applying tighter azimuth angle thresholds, or developing new criteria for edge construction. These directions will guide our next development stage.

\section{CONCLUSION}

In tracking systems that incorporate large gaseous detectors, such as drift chambers and TPC, the tracking process in the HLT typically incurs hundreds of milliseconds latencies. Conventional tracking algorithms for such detectors generally maintain stable and satisfactory performance up to an occupancy level of around 15\%. However, HLT tracking algorithm with better performance is required to meet the more challenging environments of next-generation experiments, including higher occupancy and stricter latency constraints. To simplify global tracking under complex conditions, a novel HLT tracking algorithm based on track segments is proposed. This method involves constructing a pattern bank consisting of 11 predefined patterns, optimizing the edge matrix construction using position, momentum, and time criteria, and merging stereo superlayer segments to enhance track consistency. These steps significantly reduce the number of recorded segments and the size of the edge matrix, thereby simplifying the overall global tracking complexity. In this stage, DFS is applied within each connected component to identify and deduplicate optimal paths. Simulation results demonstrate that the proposed algorithm maintains stable performance across occupancies from 5\% to 25\%, while achieving a data compression ratio of approximately 70\% even at 25\% occupancy. Meanwhile, this algorithm can handle many complex tracking cases, such as looped tracks, low-$p_t$ and ultra-low-$p_t$ₜ tracks, crossing tracks and displaced tracks. Validation against the STCF offline reconstruction algorithm confirms that the HLT tracking algorithm preserves high signal retention, without introducing noticeable negative effects on offline tracking efficiency or on the reconstruction of momentum, polar angle, and azimuth angle. These results demonstrate that the algorithm can effectively identify high-quality signal hits across a wide range of occupancy while maintaining sufficient data compression. In the next stage, the integration of neural networks is planned, and the use of segment momentum information will be optimized, with the aim of further reducing the computing latency.

\ifCLASSOPTIONcaptionsoff
  \newpage
\fi

\bibliographystyle{IEEEtran}
\bibliography{bibtex/bib/IEEEexample}

\end{document}